\documentstyle[12pt,epsfig,twoside,fleqn,espcrc1]{article}
\newcommand{\AmS}{{\protect\the\textfont2
  A\kern-.1667em\lower.5ex\hbox{M}\kern-.125emS}}
\def\bge{\begin{equation}}
\def\ene{\end{equation}}
\def\bg{\begin{eqnarray}}
\def\en{\end{eqnarray}}

\def\ubar{{\bar{u}}}
\def\dbar{{\bar{d}}}
\def\sbar{{\bar{s}}}
\def\cbar{{\bar{c}}}
\def\d0bar{{\bar{D}^0}}
\def\Dbar{{\bar{D}}}

\def\vr{\vec{r}}
\def\vx{\vec{x}}

\def\bm{\boldmath}

\title{\vspace*{-5em} Study of $\omega$-, $\eta$-, $\eta'$- and $D^-$-mesic nuclei}

\author{K. Tsushima\address{CSSM and Department of Physics
and Mathematical Physics\\ The University of Adelaide,
SA 5005, Australia}
\thanks{ADP-99-7/T352\protect\\ 
Talk given at KEK-Tanashi International Symposium on
"PHYSICS OF HADRONS AND NUCLEI",
Tokyo, December 14-17, 1998} 
}

\begin{document}

\maketitle

\begin{abstract}
Using the quark-meson coupling (QMC) model, we
investigate whether $\omega$, $\eta$, $\eta'$ and $D^-$ mesons
form meson-nucleus bound states.
Our results suggest that one should
expect to find $\eta$- and $\omega$-nucleus bound states in
all the nuclei considered. Furthermore, it is shown that the $D^-$ meson
will form quite narrow bound states with $^{208}$Pb.
\end{abstract}
\\

To study the medium modification of the
light vector ($\rho$, $\omega$ and $\phi$) meson masses is very interesting
because it is expected to provide us information concerning
chiral symmetry (restoration) in a nuclear medium.
Such experiments carried out by the CERES and HELIOS 
collaborations at the CERN/SPS~\cite{ceres}, 
and those planned at TJNAF
and GSI~\cite{jlab}, are closely related to this issue.
Recently, an alternative approach to study meson mass shifts in nuclei
was suggested by Hayano {\it et al.}~\cite{hayano} to produce
$\eta$ and $\omega$ mesons with nearly zero recoil, 
which inspired the theoretical investigations 
of $\eta$- and $\omega$-mesic nuclei~\cite{hayano2,etao}. 
The interesting point of the suggestion is that the meson 
is expected to be bound in a nucleus, if the meson feels a large enough  
attractive (Lorentz scalar) force inside the nucleus. 
We will report here the results for the mesic nuclei those studied 
using the quark-meson coupling (QMC) model~\cite{etao,dmeson}.
(See Refs.~\cite{qmc,finite1,tony} for the QMC model.)


Concerning charmed mesic nuclei~\cite{dmeson}, it is in some ways even more 
exciting, in that it promises more specific information on the
relativistic mean fields in nuclei and the nature of dynamical chiral
symmetry breaking. We focus on systems containing an anti-charm quark
and a light quark, $\bar{c}q$ $(q=u,d)$, 
which have no strong decay channels if bound.
If we assume that dynamical chiral symmetry breaking is the
same for the light quark in the charmed meson as in purely light-quark
systems, we expect the same coupling constant,
$g^q_\sigma$, in QMC.
In the absence of any strong interaction, the $D^-$ will form
atomic states, bound by the Coulomb potential.
The resulting binding for, say, the 1s level in $^{208}$Pb
is between ten and thirty MeV and should provide a very clear
experimental signature. On the other hand,
although we expect the D-meson (systems of $\bar{q}c$) will
form deeply bound $D$-nucleus states, they will also couple
strongly to open channels such as $D N \rightarrow B_c (\pi's)$, with
$B_c$ a charmed baryon.
Unfortunately, because our present knowledge does not permit
an accurate calculation of the $D$-meson widths in a nucleus,
results for the $D$-mesic nuclei may not give useful information
for experimenters.

\begin{figure}[htb]
\begin{minipage}[t]{75mm}
\epsfig{file=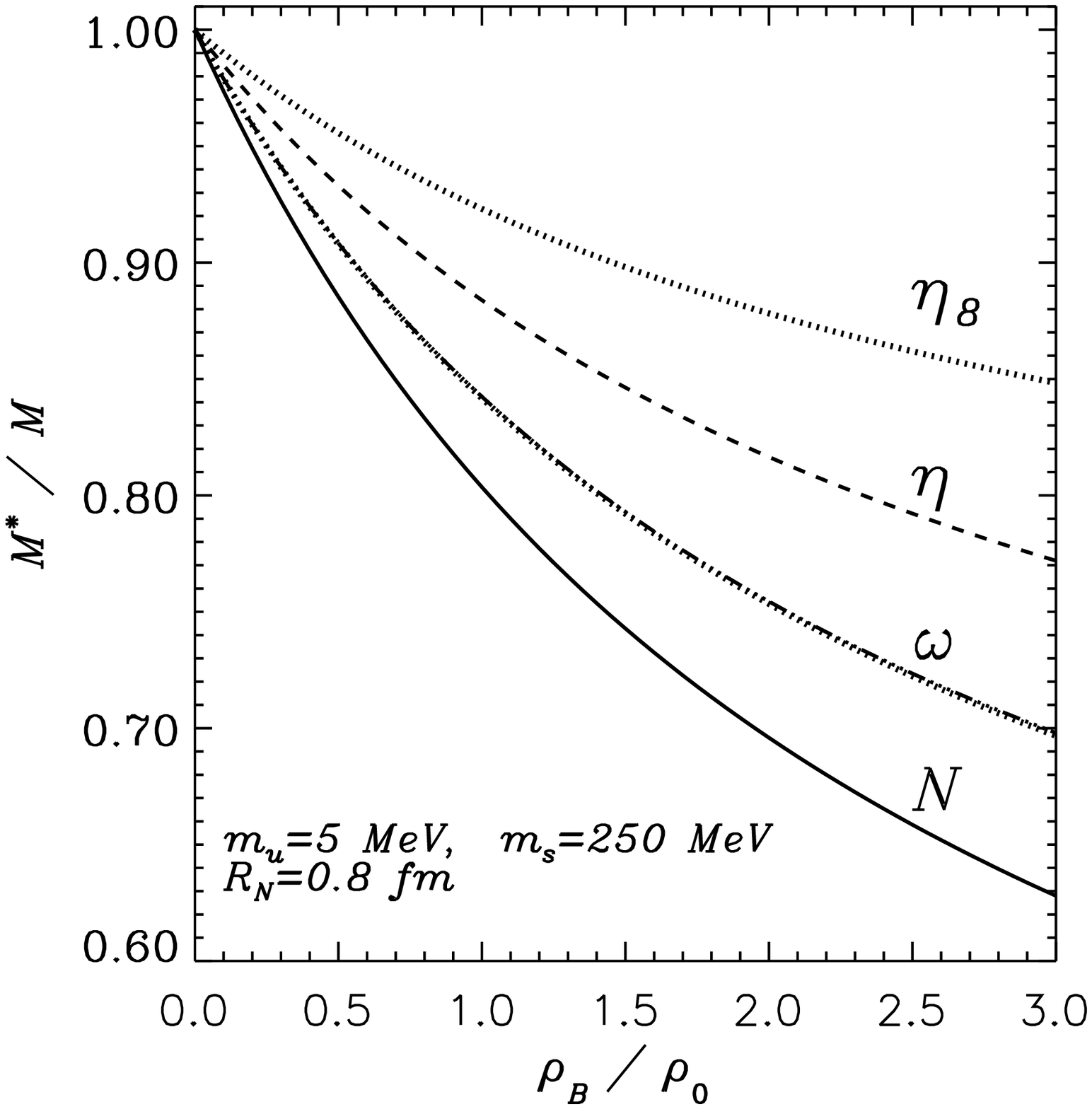,height=6cm}
\vspace{-3em}
\caption{Effective masses in symmetric nuclear matter.
($\rho_0$ = 0.15 fm$^{-3}$.)}
\label{etaomassmatter}
\end{minipage}
\hspace{\fill}
\begin{minipage}[t]{75mm}
\epsfig{file=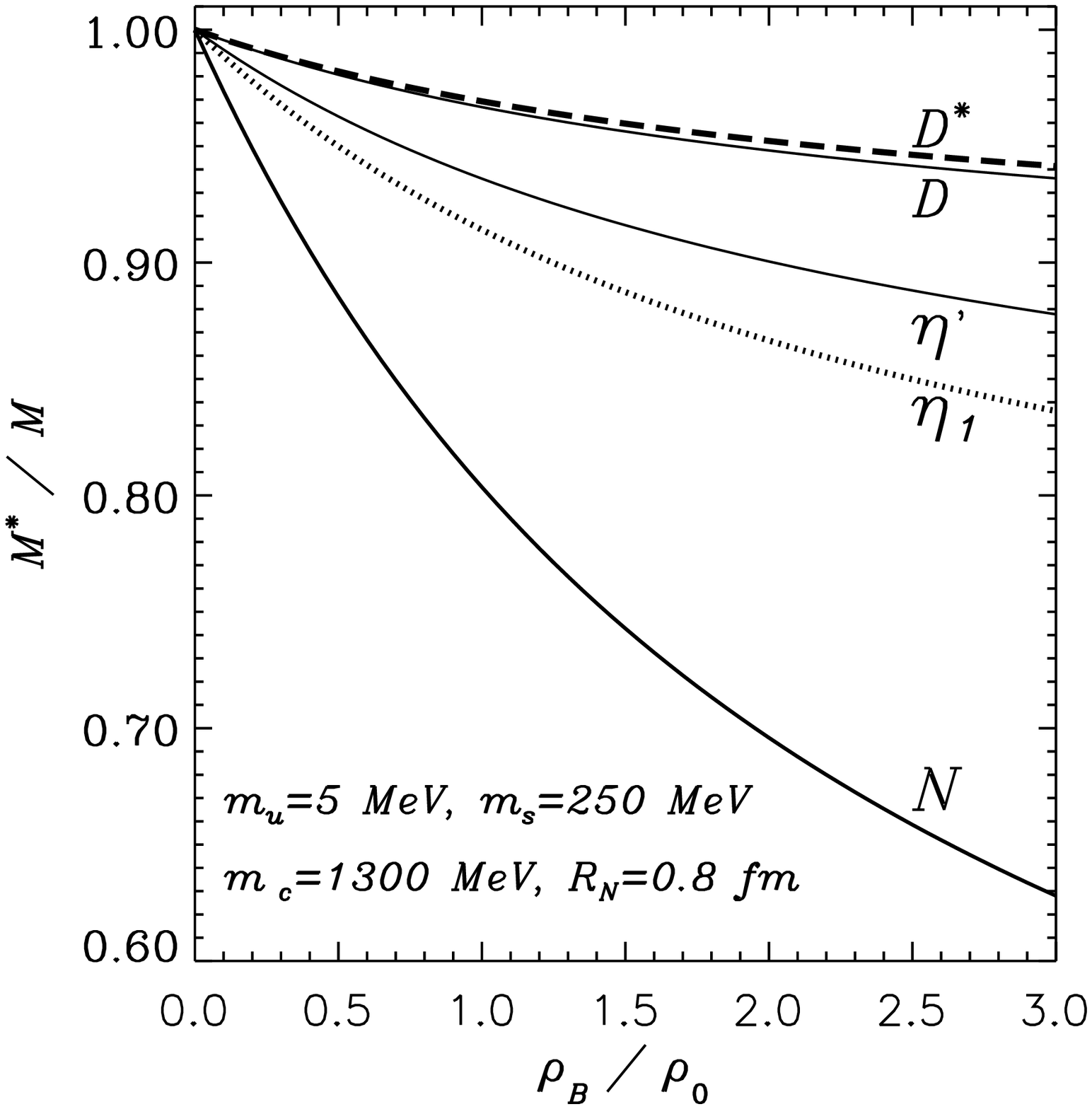,height=6cm}
\vspace{-3em}
\caption{See the caption of Fig.~\protect\ref{etaomassmatter}.}
\label{detapmatter}
\end{minipage}
\vspace{-2em}
\end{figure}

In Figs.~\ref{etaomassmatter} and~\ref{detapmatter} 
we show the mass shifts of the mesons  
in symmetric nuclear matter~\cite{etao,dmeson}.
The masses for the physical $\omega$, $\eta$ and $\eta'$ are calculated  
using the octet and singlet states with the mixing angles,
$\theta_P=-10^\circ$ for ($\eta,\eta'$), and
$\theta_V=39^\circ$ for ($\phi,\omega$)~\cite{pdata}.
The masses for the octet and singlet states without the 
mixing are also shown (the dotted lines).
%
%
%

%
\begin{figure}[htb]
\begin{minipage}[t]{75mm}
\epsfig{file=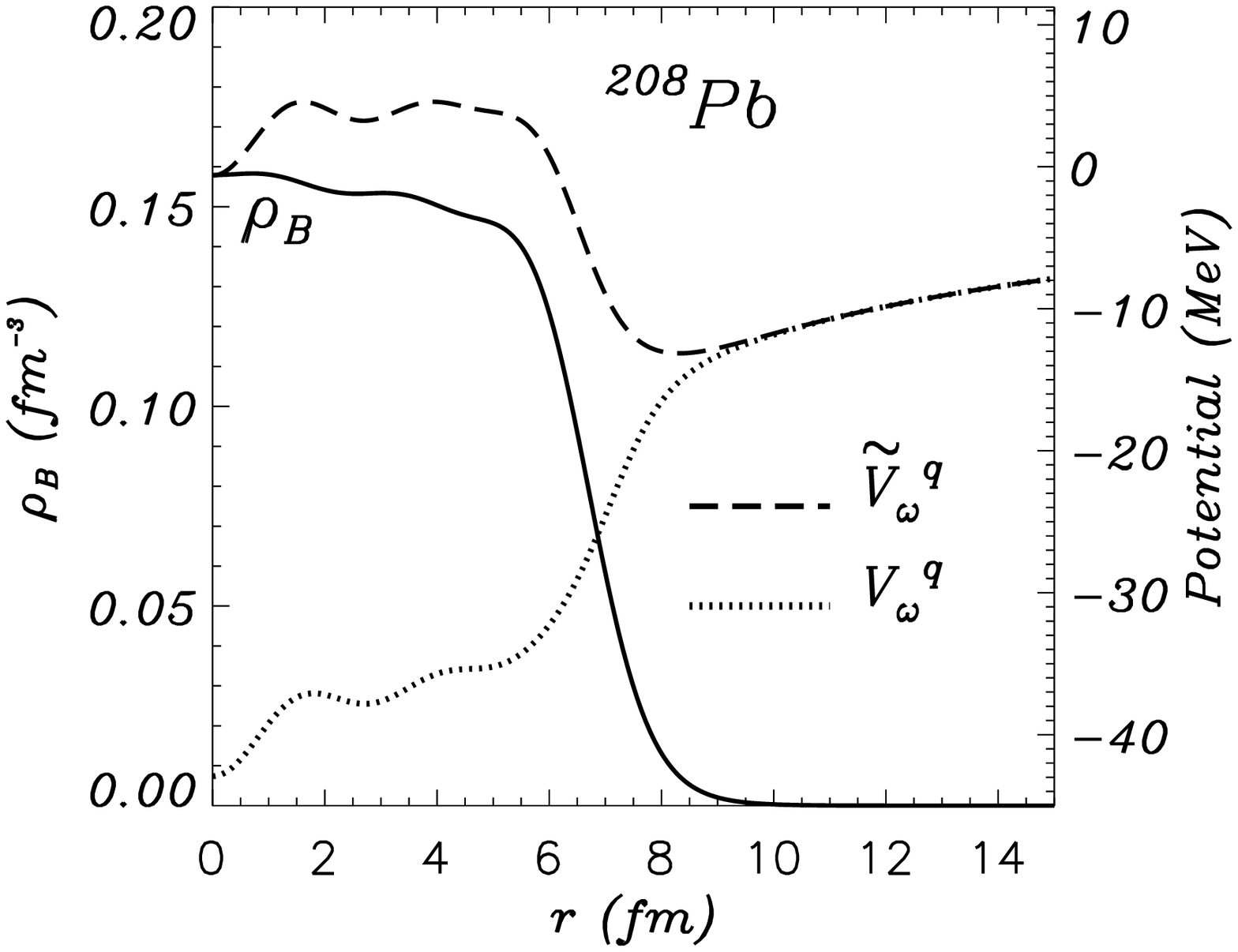,height=5.0cm}
\vspace{-3em}
\caption{Total potential for the $D^-$.
}
\label{dmespot}
\end{minipage}
\hspace{\fill}
\begin{minipage}[t]{75mm}
\epsfig{file=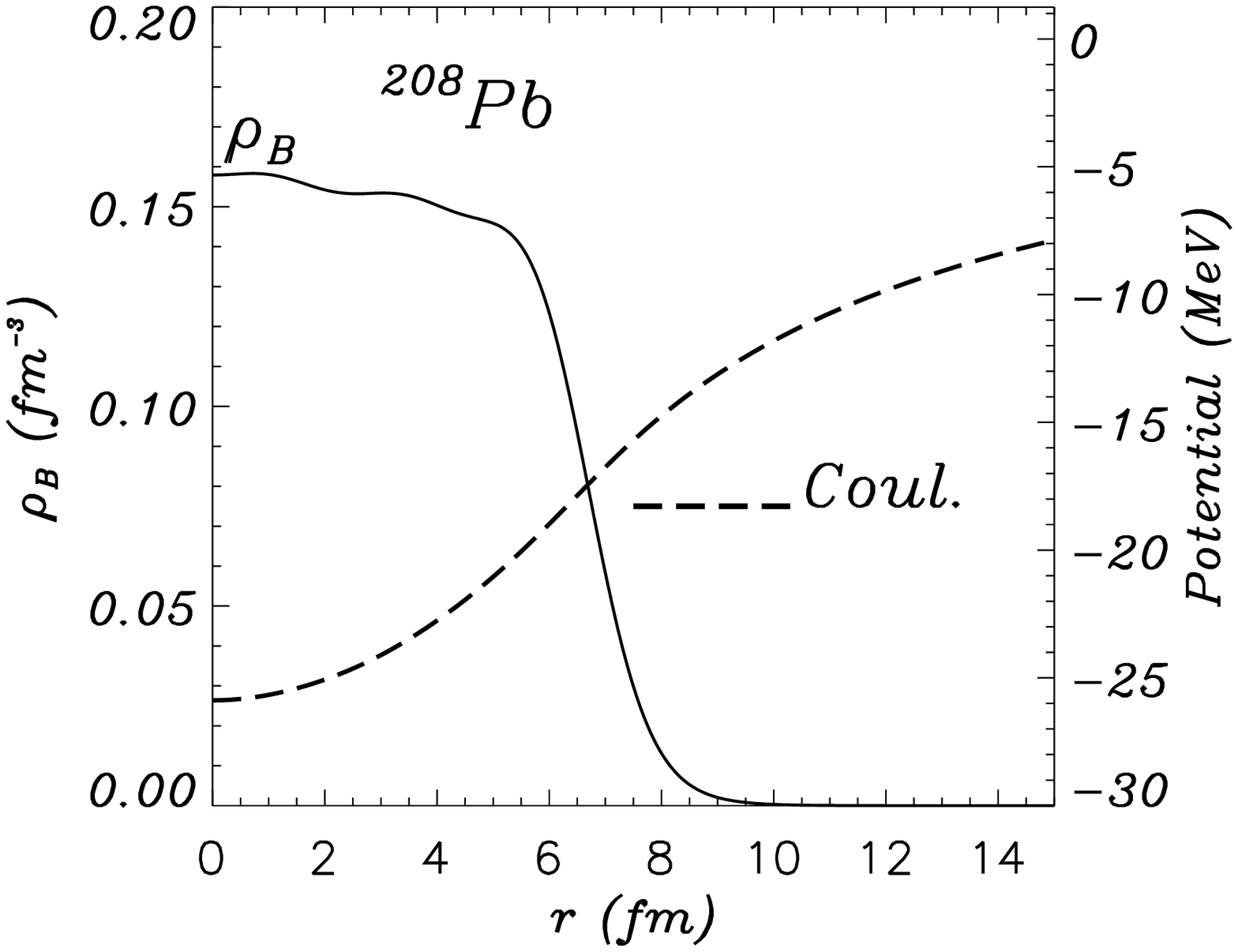,height=5.0cm}
\vspace{-3em}
\caption{Coulomb potential for the $D^-$.}
\label{dcoulombpb}
\end{minipage}
\vspace{-1em}
\end{figure}
%

%
%
At position $\vr$ in a nucleus
the Dirac equations for the quarks and antiquarks in the
meson bags ($|\vx - \vr| \le$ bag radius)
are given by~\cite{etao,dmeson,kaon}:
\bg
\left[ i \gamma \cdot \partial_x - (m_q - V^q_\sigma(\vr))
\mp \gamma^0
\left( V^q_\omega(\vr) + \frac{1}{2} V^q_\rho(\vr) \right) \right]
\left(\begin{array}{c} \psi_u(x)\\ \psi_\ubar(x)\\ \end{array}\right)
 &=& 0,
\label{diracu}
\\
\left[ i \gamma \cdot \partial_x - (m_q - V^q_\sigma(\vr))
\mp \gamma^0
\left( V^q_\omega(\vr) - \frac{1}{2} V^q_\rho(\vr) \right) \right]
\left(\begin{array}{c} \psi_d(x)\\ \psi_\dbar(x)\\ \end{array} \right)
 &=& 0,
\label{diracd}
\\
\left[ i \gamma \cdot \partial_x - m_{s,c} \right]
\psi_{s,c} (x)\,\, ({\rm or}\,\, \psi_{\sbar,\cbar}(x)) &=& 0,
\label{diracsc}
\en
where the mean-field potentials for a bag centered at position $\vr$ in
the nucleus, 
are calculated self-consistently in QMC by solving
Eqs.~(23) -- (30) of Ref.~\cite{finite1}.

%

Then, the meson masses
in the nucleus are calculated by~\cite{etao,dmeson}:
%
\bg
m_{\eta,\omega}^*(\vr) &=& \frac{2 [a_{P,V}^2\Omega_q^*(\vr)
+ b_{P,V}^2\Omega_s(\vr)] - z_{\eta,\omega}}{R_{\eta,\omega}^*}
+ {4\over 3}\pi R_{\eta,\omega}^{* 3} B,
\qquad ({\rm for}\, \eta',\, a_P \leftrightarrow b_P),
\label{metao}\\
m_{D}^*(\vr) &=& \frac{\Omega_q^*(\vr)
+ \Omega_c(\vr) - z_{D}}{R_{D}^*}
+ {4\over 3}\pi R_{D}^{* 3} B,
\label{md}\\
& &\left.\frac{\partial m_j^*(\vr)}
{\partial R_j}\right|_{R_j = R_j^*} = 0, \quad\quad
(j = \omega,\eta,\eta',D),
\label{equil}\\
%
%
a_{P,V} &\equiv& \sqrt{1/3} \cos\theta_{P,V}
- \sqrt{2/3} \sin\theta_{P,V},\quad
b_{P,V} \equiv \sqrt{2/3} \cos\theta_{P,V}
+ \sqrt{1/3} \sin\theta_{P,V},\qquad
\label{abpv}
\en
where $\Omega_q^*(\vr) = \sqrt{x_q^2 + (R_j^* m_q^*)^2}$, with
$m_q^* = m_q - g^q_\sigma \sigma(\vr)$ and
$\Omega_{s,c}(\vr) = \sqrt{x_{s,c}^2 + (R_j^* m_{s,c})^2}$.

In this study we chose the values,
$(m_q, m_s, m_c) = (5, 250, 1300)$ MeV for the
current quark masses, and $R_N = 0.8$
fm for the bag radius of the nucleon in free space.
(See Ref.~\cite{finite1} for the other parameters.)
We stress that exactly the same coupling constants
in QMC, $g^q_\sigma$, $g^q_\omega$ and
$g^q_\rho$, are used for the light quarks in the mesons as in the
nucleon. However, in studies of the kaon system, we found that it was
phenomenologically necessary to increase the strength of the vector
coupling, $g^q_\omega$, in the $K^+$
($g^q_\omega \to 1.4^2 g^q_\omega$, i.e., 
$V^q_\omega(r) \rightarrow \tilde{V}^q_\omega(r) = 1.4^2 V^q_\omega(r)$)
in order to reproduce the empirically extracted $K^+$-nucleus
interaction~\cite{kaon}.
This may be ascribed to the fact that the kaon is a pseudo-Goldstone boson
and expected to be difficult to treat properly with the usual bag model.
Thus, we show
results for the $\Dbar$ bound state energies with both choices for
the vector potential, $V^q_\omega(r)$ and $\tilde{V}^q_\omega(r)$. 

In Fig.~\ref{dmespot} we show the calculated
potentials for the $D^-$.
The left panel shows the {\it naive} sum of the potentials,
$(m^*_{D^-}(r) - m_{D^-}) + [V^q_\omega(r)$ or $\tilde{V}^q_\omega(r)]
+ \frac{1}{2} V^q_\rho(r) - A(r)$ (the dotted or dashed line). 
The right panel shows the Coulomb potential.
One expects the existence of the $_{D^-}^{208}$Pb states
just from inspection of the sum of the potentials, 
because the $D^-$ is heavy and may be described well 
in the nonrelativistic Schr\"{o}dinger equation.


\begin{table}[htb]
\vspace{-2em}
\begin{center}
\caption{
$\eta$, $\omega$ and $\eta'$ bound state energies (in MeV),
$E_j = Re (E^*_j - m_j)\,(j=\eta,\omega,\eta')$,
where all widths for the $\eta'$ are set to zero.
The eigenenergies are given by,
$E^*_j = E_j + m_j - i \Gamma_j/2$.
}
\label{etaoenergy}
\begin{tabular}[t]{lcccccl}
\hline 
& &$\bm \gamma_\eta=0.5$ & &$\bm \gamma_\omega$=0.2
& &$\bm \gamma_{\eta'}=0$ \\
\hline 
& &$E_\eta$ &$\Gamma_\eta$ &$E_\omega$ &$\Gamma_\omega$
&$E_{\eta'}$ \\
\hline
$^{6}_j$He &1s &-10.7&14.5 &-55.6&24.7 &* (not calculated)\\
%
$^{11}_j$B &1s &-24.5&22.8 &-80.8&28.8 &* \\
%
$^{26}_j$Mg &1s &-38.8&28.5 &-99.7&31.1 &* \\
            &1p &-17.8&23.1 &-78.5&29.4 &* \\
            &2s & --- & --- &-42.8&24.8 &* \\
%
$^{16}_j$O &1s &-32.6&26.7 &-93.4&30.6 &-41.3 \\
           &1p &-7.72&18.3 &-64.7&27.8 &-22.8 \\
%
$^{40}_j$Ca &1s &-46.0&31.7 &-111&33.1  &-51.8 \\
            &1p &-26.8&26.8 &-90.8&31.0 &-38.5 \\
            &2s &-4.61&17.7 &-65.5&28.9 &-21.9 \\
%
$^{90}_j$Zr &1s &-52.9&33.2 &-117&33.4  &-56.0 \\
            &1p &-40.0&30.5 &-105&32.3  &-47.7 \\
            &2s &-21.7&26.1 &-86.4&30.7 &-35.4 \\
%
$^{208}_j$Pb &1s &-56.3&33.2 &-118&33.1 &-57.5 \\
             &1p &-48.3&31.8 &-111&32.5 &-52.6 \\
             &2s &-35.9&29.6 &-100&31.7 &-44.9 \\
\hline 
\end{tabular}
\vspace{-2em}
\end{center}
\end{table}
%
%
\begin{table}[htb]
\vspace{-2em}
\begin{center}
\caption{
$D^-$, $\d0bar$ and $D^0$ bound state energies (in MeV).
The widths are all set to zero. }
\label{denergy}
\begin{tabular}[t]{lcccccc}
\hline
state  &$D^- (\tilde{V}^q_\omega)$ &$D^- (V^q_\omega)$
&$D^- (V^q_\omega$, no Coulomb) &$\d0bar (\tilde{V}^q_\omega)$
&$\d0bar (V^q_\omega)$ &$D^0 (V^q_\omega)$ \\
\hline
                         1s &-10.6 &-35.2 &-11.2 &unbound &-25.4 &-96.2\\
                         1p &-10.2 &-32.1 &-10.0 &unbound &-23.1 &-93.0\\
                         2s & -7.7 &-30.0 & -6.6 &unbound &-19.7 &-88.5\\
\hline
\end{tabular}
\end{center}
\vspace{-2em}
\end{table}
%

To calculate the bound state energies for the
mesons with the situation of almost zero momenta, we may 
solve the following form of the Klein-Gordon equation~\cite{etao,dmeson}:
%
\bg
[ \nabla^2 &+& (E^*_j - V^j_v(r))^2- \tilde{m}^{*2}_j(r) ]\,
\phi_j(\vr) = 0,\qquad (j=\omega,\eta,\eta',D),
\label{kgequation}
\\
\tilde{m}^*_j(r) &\equiv& m^*_j(r) - \frac{i}{2}
\left[ (m_j - m^*_j(r)) \gamma_j + \Gamma_j^0 \right]\,
\equiv\, m^*_j(r) - \frac{i}{2} \Gamma^*_j (r),
\label{width}
\en
where $E^*_j$ is the total energy of the meson,
$V^j_v(r)$, $m_j$ and $\Gamma_j^0$
are the sum of the vector and Coulomb potentials,
the corresponding masses and widths in free space, and
$\gamma_j$ are treated as phenomenological
parameters to describe the in-medium meson widths,
$\Gamma^*_j(r)$~\cite{etao,dmeson}.
%
%
We show the bound state energies calculated for
$\gamma_\eta = 0.5$ and $\gamma_\omega = 0.2$,
which are expected to correspond best with experiment
according to the estimates in Refs.~\cite{hayano,fri}.
For the $D^-$ and $\d0bar$, the widths are set to zero which is exact,
whereas those for the $\eta'$ and $D^0$ do not make sense.
%
%
The calculated bound state energies
are listed in Tables~\ref{etaoenergy} and~\ref{denergy}.


Our results suggest that $\eta$ and $\omega$ mesons should be
bound in all the nuclei considered. Furthermore, the $D^-$ meson
should be bound in $^{208}$Pb, due to two 
different mechanisms, namely, the scalar and attractive
$\sigma$ mean field for the case of $V^q_\omega(r)$ even without the Coulomb
force, or solely due to the Coulomb force for the case of
$\tilde{V}^q_\omega(r) = 1.4^2 V^q_\omega(r)$.
The existence of any bound states at all would give us important
information concerning the role of the Lorentz scalar $\sigma$ field,
and hence dynamical symmetry breaking.

\vspace{1em}
\noindent{\bf Acknowledgment:} 
The author would like to thank D.H. Lu, K. Saito and
A.W. Thomas for exciting collaborations.
This work was supported by the Australian Research Council.
%

%

\begin{thebibliography}{9}
%
\bibitem{ceres}P. Wurm for the CERES collaboration,
Nucl. Phys. A 590 (1995) 103c;
M. Masera for the HELIOS collaboration, Nucl. Phys.
A 590 (1995) 93c.
%
%
\bibitem{jlab}M. Kossov et al., TJNAF proposal PR-94-002 (1994);
P.Y. Bertin and P.A.M. Guichon, Phys. Rev. C 42 (1990) 1133;
HADES proposal,~http://piggy.physik.uni-giessen.de/hades/;
G.J. Lolos et al., Phys. Rev. Lett. 80 (1998) 241.
%
\bibitem{hayano}R.S. Hayano et al., proposal for
GSI/SIS, September, 1997.
%
\bibitem{hayano2}R.S. Hayano, S. Hirenzaki and A. Gillitzer, nucl-th/9806012;
F. Klingl, T. Waas and W. Weise, hep-ph/9810312.
%
\bibitem{etao}K. Tsushima, D.H. Lu, A.W. Thomas, K. Saito,
Phys. Lett. B 443 (1998) 26.
%
\bibitem{dmeson}K. Tsushima, D.H. Lu, A.W. Thomas, K. Saito and R.H. Landau,
ADP-98-48/T317, OSUNT98-13, nucl-th/9810016;
K. Tsushima, nucl-th/9811063.
%
\bibitem{qmc}P.A.M. Guichon, Phys. Lett. B 200, 235 (1988);
P.A.M. Guichon, K. Saito, E. Rodionov and A.W. Thomas,
Nucl. Phys. A 601 (1996) 349.
%
\bibitem{finite1}K. Saito, K. Tsushima and A.W. Thomas,
Nucl. Phys. A 609 (1996) 339.
%
\bibitem{tony}K. Saito, K. Tsushima and A.W. Thomas,
Phys. Rev. C 55 (1997) 2637;
K. Tsushima et al., 
Nucl. Phys. A 630 (1998) 691;
A.W. Thomas, nucl-th/9807027.
%
\bibitem{pdata}Review of Particle Physics, Phys. Rev. D 54, 1 (1996).
%
\bibitem{kaon}K. Tsushima, K. Saito, A.W. Thomas and S.W. Wright,
Phys. Lett. B 429, 239 (1998);
{\it ibid} (E) Phys. Lett. B 436 (1998) 453.
%
%
\bibitem{fri}B. Friman,  nucl-th/9801053;
F. Klingl and W. Weise,  hep-ph/9802211.
%
%
\end{thebibliography}
\end{document}